\documentclass[
reprint, 
superscriptaddress, 
amsmath, 
amssymb, 
aps, 
prb
]{revtex4-1}

\usepackage{graphicx}% Include figure files
\usepackage{dcolumn}% Align table columns on decimal point
\usepackage{bm}
\usepackage{color}
\begin{document}
\preprint{APS/123-QED}
%\preprint{}

\title{Origin of magnetovolume effect in a cobaltite}


\author{Ping Miao}
\email{miao@post.kek.jp}
\affiliation{Institute of Materials Structure Science, High Energy Accelerator Research Organization (KEK), Tokai 319-1106, Japan}

\author{Zhijian Tan}
\affiliation{Institute of Materials Structure Science, High Energy Accelerator Research Organization (KEK), Tokai 319-1106, Japan}
\affiliation{Sokendai (The Graduate University for Advanced Studies), Tokai 319-1106, Japan}

\author{Sanghyun Lee}
\affiliation{Institute of Materials Structure Science, High Energy Accelerator Research Organization (KEK), Tokai 319-1106, Japan}

\author{Yoshihisa Ishikawa}
\affiliation{Institute of Materials Structure Science, High Energy Accelerator Research Organization (KEK), Tokai 319-1106, Japan}

\author{Shuki Torii}
\affiliation{Institute of Materials Structure Science, High Energy Accelerator Research Organization (KEK), Tokai 319-1106, Japan}

\author{Masao Yonemura}
\affiliation{Institute of Materials Structure Science, High Energy Accelerator Research Organization (KEK), Tokai 319-1106, Japan}
\affiliation{Sokendai (The Graduate University for Advanced Studies), Tokai 319-1106, Japan}

\author{Akihiro Koda}
\affiliation{Institute of Materials Structure Science, High Energy Accelerator Research Organization (KEK), Tokai 319-1106, Japan}
\affiliation{Sokendai (The Graduate University for Advanced Studies), Tokai 319-1106, Japan}

\author{Kazuki Komatsu}
\affiliation{Geochemical Research Center, Graduate School of Science, The University of Tokyo, Tokyo 113-0033, Japan}

\author{Shinichi Machida}
\affiliation{Neutron Science and Technology Center, CROSS, Tokai, Ibaraki 319-1106, Japan}

\author{Asami Sano-Furukawa}
\affiliation{J-PARC Center, Japan Atomic Energy Agency (JAEA), Tokai, Ibaraki 319-1195, Japan}

\author{Takanori Hattori}
\affiliation{J-PARC Center, Japan Atomic Energy Agency (JAEA), Tokai, Ibaraki 319-1195, Japan}

\author{Xiaohuan Lin}
\affiliation{Center for High Pressure Science and Technology Advanced Research, Beijing, 100094, China}
\affiliation{College of Chemistry and Molecular Engineering, Peking University, Beijing 100871, China}

\author{Kuo Li}
\affiliation{Center for High Pressure Science and Technology Advanced Research, Beijing, 100094, China}

\author{Takashi Mochiku}
\affiliation{National Institute for Materials Science (NIMS), Tsukuba, Ibaraki 305-0047, Japan}

\author{Ryosuke Kikuchi}
\affiliation{Department of Physics, College of Humanities and Sciences, Nihon University, Tokyo, 156-8550, Japan}

\author{Chizuru Kawashima}
\affiliation{Department of Physics, College of Humanities and Sciences, Nihon University, Tokyo, 156-8550, Japan}

\author{Hiroki Takahashi}
\affiliation{Department of Physics, College of Humanities and Sciences, Nihon University, Tokyo, 156-8550, Japan}

\author{Qingzhen Huang}
\affiliation{NIST Center for Neutron Research, National Institute of Standards and Technology, 100 Bureau Drive, Gaithersburg, MD 20899, United States}

\author{Shinichi Itoh}
\affiliation{Institute of Materials Structure Science, High Energy Accelerator Research Organization (KEK), Tokai 319-1106, Japan}
\affiliation{Sokendai (The Graduate University for Advanced Studies), Tokai 319-1106, Japan}

\author{Ryosuke Kadono}
\affiliation{Institute of Materials Structure Science, High Energy Accelerator Research Organization (KEK), Tokai 319-1106, Japan}
\affiliation{Sokendai (The Graduate University for Advanced Studies), Tokai 319-1106, Japan}

\author{Yingxia Wang}
\affiliation{College of Chemistry and Molecular Engineering, Peking University, Beijing 100871, China}

\author{Feng Pan}
\affiliation{School of Advanced Materials, Peking University, Shenzhen Graduate School, Shenzhen 518055, China}

\author{Kunihiko Yamauchi}
\email{kunihiko@sanken.osaka-u.ac.jp}
\affiliation{ISIR-SANKEN, Osaka University, Ibaraki, Osaka 567-0047, Japan}

\author{Takashi Kamiyama}
\email{takashi.kamiyama@kek.jp}
\affiliation{Institute of Materials Structure Science, High Energy Accelerator Research Organization (KEK), Tokai 319-1106, Japan}
\affiliation{Sokendai (The Graduate University for Advanced Studies), Tokai 319-1106, Japan}


\date{\today}

\begin{abstract}

The layered perovskite PrBaCo$_2$O$_{5.5}$ demonstrates a strong negative thermal expansion (NTE) which holds potential for being fabricated into composites with zero thermal expansion. The NTE was found to be intimately associated with the spontaneous magnetic ordering, known as magnetovolume effect (MVE). Here we report with compelling evidences that the continuous-like MVE in PrBaCo$_2$O$_{5.5}$ is intrinsically of discontinuous character, originating from an magnetoelectric transition from an antiferromagnetic insulating large-volume (AFILV) phase to a ferromagnetic metallic small-volume (FMSV) phase. Furthermore, the magnetoelectric effect (ME) shows high sensitivity to multiple external stimuli such as temperature, carrier doping, hydrostatic pressure, magnetic field etc. In contrast to the well-known ME such as colossal magnetoresistance and multiferroic effect which involve symmetry breaking of crystal structure, the ME in the cobaltite is purely isostructural.  Our discovery provides a new pathway to realizing the ME as well as the NTE, which may find applications in new techniques. 

\end{abstract}

%\keywords{}

\maketitle

\section{Introduction}

The ``Invar effect" originates from the discovery by Guillaume\cite{Guillaume_1897} in 1897 that the Fe$_{65}$Ni$_{35}$ alloy undergoes almost zero thermal expansion in a wide range of temperature. Since then, various alloys with very low thermal expansion coefficient, known as Invar alloys, have been developed and applied to fields where dimensional stability is required, ranging from precision instruments such as telescopes, standard rulers, timing devices etc., to large structural components like railroad tracks, bridges, liquefied natural gas containers and so on\cite{Wasserman_1990}. Although it is generally agreed that the normal positive thermal expansion from phonons in Invar alloys is compensated by a negative contribution arising from spontaneous magnetic ordering, which is known as the magnetovolume effect (MVE)\cite{Wasserman_1990, Nakamura_1976}.  Although diverse theoretical models from different perspectives have been proposed to explain the profound MVE\cite{Weiss_1963, Moriya_1980, Lagarec_2001, Abrikosov_2009, Johansson_2007, Crisan_2002, Kaspar_1981, Khmelevskyi_2003, Kondorsky_1960, Rancourt_1996, Johansson_1999, Hasegawa_1983, Liot_2008, Staunton_1987}, there is still no consensus on the microscopic origin. One of the difficulties in experimentally justifying the theories lies in the fact the alloys contain chemical inhomogeneities that hinder the detection of intrinsic electromagnetic inhomogeneities. 

\begin{figure*}
\includegraphics[width=0.7\textwidth]{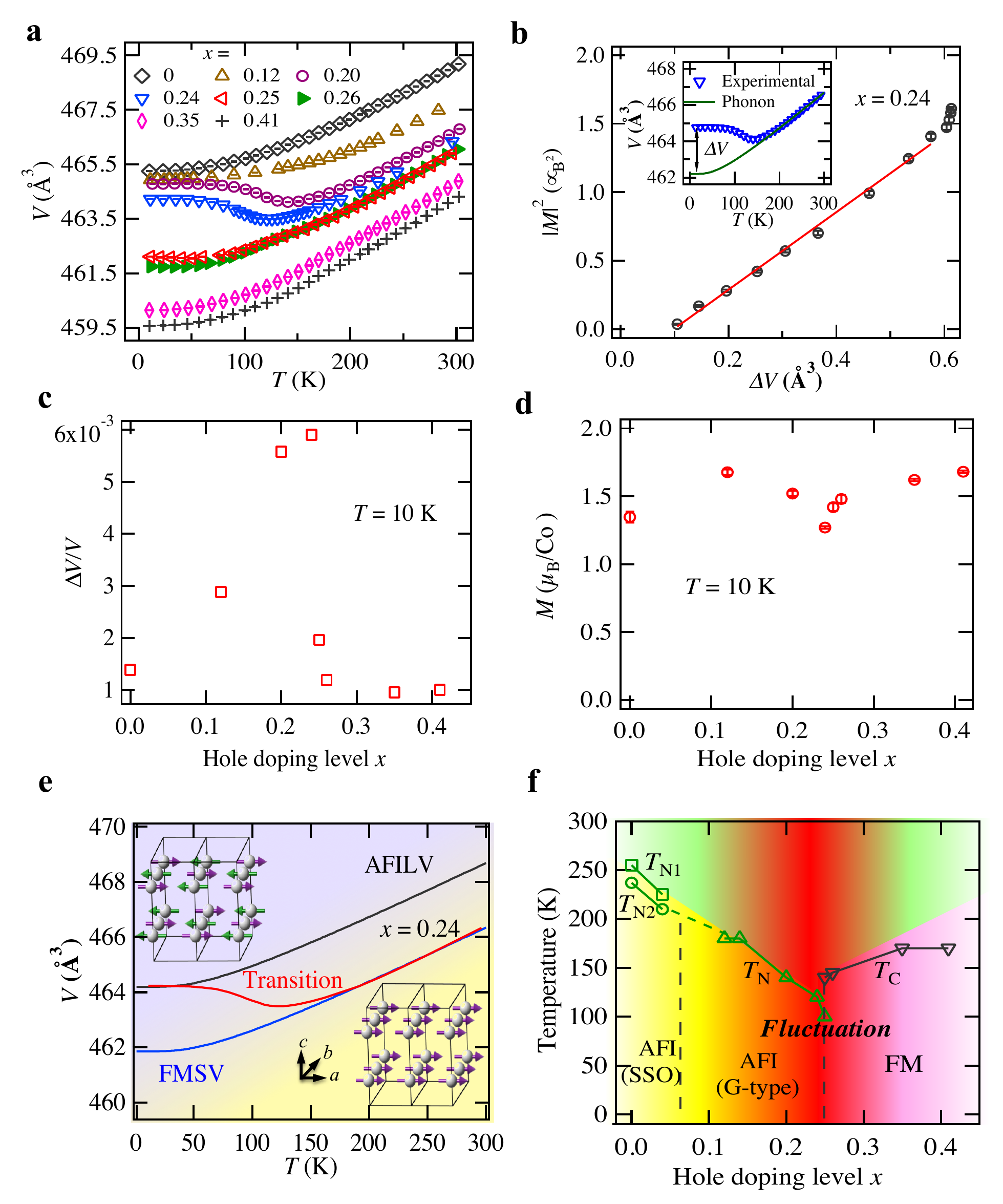}
\caption{\label{fig1} \textbf{$\vert$ Unit cell volume (a), magnetovolume (b, c), Co-ion magnetic moments (d), transition model (e) and phase diagram (f) of PrBaCo$_\textbf{2}$O$_\textbf{5.5+x}$. $\vert$}  All the results are derived from analysis of high-resolution NPD data. (a), Volume of sub unit cell ($2a_p \times 2a_p \times 2a_p$) as a function of temperature for various hole-doping fraction $x$ which shows that anomalous thermal expansion is enhanced with increasing x until maximum at $x$ = 0.24 and drops with further hole doping. (b), Square of magnetic moment $\lvert M \rvert ^2$ as a function of volumetric order parameter $\Delta V$ for $x$ = 0.24. $M$ is determined from Rietveld refinement using a G-type antiferromagnetic structure model. $\Delta V$ is extracted by subtracting the phonon contribution, which is calculated on the basis of Debye$-$Gruneisen model as shown in the inset. The solid line is the linear fit to the data. The $M$ arises continuously from zero so that the MVE looks like the second-order (continuous) phase transition. (c), Spontaneous magetovolume contribution at 10 K as a function of hole-doping fraction $x$. (d), Magnetic moment of single $Co$ ion at $T$ = 10 K as a function of hole doping level $x$, which is determined from Rietveld refinement. The residual values R$_{wp}$ is both below 10\% and R$_{M}$ below 20\%. For the spin-state ordered (SSO) phase at $x$ = 0, where different crystallographic sites of $Co$ bear different magnetic moments, the magnetic moment is averaged over all $Co$ sites. For the coexisting long-range ordered ferromagnetic and antiferromagnetic phases at $x$ = 0.25, the magnetic moment is averaged over both phases. (e), The magnetoelectric transition model for NTE of $x$ = 0.24, $i.e.$, the transition from an AFILV phase with antiferromagnetic (G-type) structure of $Co$ spins (the upper-left drawing) to a FMSV phase with the ferromagnetic structure (the lower-right drawing). (f), Phase diagram for PrBaCo$_2$O$_{5.5+x}$, summarized from our previous21 and present NPD experiments. AFI(SSO), AFI(G-type), and FM denote the antiferromagnetic insulating (spin-state ordering), antiferromagnetic insulating (G-type), and ferromagnetic metallic states, respectively. $T_{N1}$ (green squares) and $T_{N2}$ (green circles) $T_{N}$ (green triangles), and $T_{C}$ (black triangles) represent for the transition temperatures for these long-range ordered magnetic structures. Giant phase fluctuations occur near the boundary between FM and AFI phases, as highlighted by the red shadows. The coexistence of long-range ordered AFILV and FMSV at $x$ = 0.25 were evidenced from peak splitting and sharp F and AF Bragg peaks at low temperatures, so that both $T_{N}$ and $T_{C}$ are marked for $x$ = 0.25 (on the dashed line).}
\end{figure*}

Oxides have been proved to be chemically homogeneous platforms for investigating electromagnetic phase separations. For example, multiple experimental evidences have demonstrated the coexistence of the antiferromagnetic Mott-insulating phase and superconducting phase in cuprates\cite{Lee_2006}, as well as the competition between the antiferromagnetic charge-ordered/orbital-ordered insulating phase and ferromagnetic charge-disordered/orbital-disordered metallic phase in colossal magnetoresistive (CMR) manganites\cite{Tokura_2006, Dagotto_2001}. Therefore, we investigated the MVE in a cobaltite with layered perovskite structure, PrBaCo$_2$O$_{5.5+x}$, which crystallizes into the $Pmmm (a_p \times 2a_p \times 2a_p)$ structure at hole doping level $0.06 < x \le0.41$. As shown in Fig.~\ref{fig1}a and Fig.~\ref{fig1}b, the cobaltite exhibits strong negative thermal expansion (NTE) upon holes doping and the continuous-like MVE is evidenced by the linear correlation between the volumetric order parameter $\Delta V$ and the square of magnetic moment $M$21. In analogy with the Invar alloys, the doping dependence of MVE also shows a peak centered at hole doping level $x \approx 0.24$ (Fig.~\ref{fig1}c)\cite{Nakamura_1976, Miao_AM_2017}. From this perspective, our study on PrBaCo$_2$O$_{5.5+x}$. may shed light on the controversial origin for MVE in Invar alloys. 

The experimental evidences and theoretical calculations reported here reveals that the ground-state near the boundary in the phase diagram, at $x \approx 0.24$ , has two separate energy minima, i.e., an antiferromagnetic insulating large-volume (AFILV) phase and a ferromagnetic metallic small-volume (FMSV) phase. As shown in Fig.~\ref{fig1}e, both phases undergo normal positive thermal expansion at finite temperatures, following the Debye-Gruneisen model that accounts for the phonon-induced thermal expansion, and the NTE comes from temperature-induced transition from AFILV phase to FMSV phase. Our discovery unveils a new mechanism for the MVE, which might shed light on the microscopic origin for Invar effect. 

The magnetoelectric transition is intimately connected with strong phase fluctuations near the boundary (Fig.~\ref{fig1}f), where the AFILV and FMSV phases intensely compete with each other. As a result, moderate external stimuli in addition to temperature ($T$), such as carrier doping ($x$), hydrostatic pressure ($P$), magnetic field ($H$) etc., can also induce the conversion between the two phases, triggering giant multiple responses. To be noted, there is no symmetry breaking of the crystal structure in the process of the AFILV$-$FMSV phase transition, which is different from the well-known magnetoelectric effect (ME) in bulk materials like CMR and multiferroic effect. As we know, the charge/orbital order in the insulating phase melts upon transitioning into the metallic phase in CMR materials\cite{Tokura_2006, Cheong_2007} while the inversion symmetry breaks so as to induce ferroelectricity in multiferroics\cite{Khomskii_2009, Dagotto_2001}.  The isostructural AFILV$-$FMSV phase transition in the cobaltite opens another way of generating ME in bulk materials, which holds substantial potential for new industrial applications. 

%Figure1

\section{Method}

\subsection{Sample preparation}
PrBaCo$_2$O$_{5.5+x}$ polycrystalline samples were synthesized by the solid-state reaction method with a combined EDTA$-$citrate complex sol$-$gel process and the oxygen content was controlled by annealing the as-prepared samples in different gas atmospheres. \cite{Miao_AM_2017, Miao_PRB_2017} High crystallinity of the samples was identified by high-resolution NPD experiments.

\subsection{Neutron powder diffraction (NPD)}
The high-resolution NPD measurements were performed using SuperHRPD\cite{Torii_2011} at Japan Proton Accelerator Research Complex (J-PARC). The averaged resolution for the back-scattering detector complex (165$^{\circ}$ $<$ 2$\theta$ $<$ 175$^{\circ})$ is $\frac{\Delta d}{d}$ = 0.09\%. The samples were mounted in a top-loading closed cycle refrigerator with base temperature $T$ = 10 K.

The magnetic-field NPD measurements were also carried out at SuperHRPD by implementing the Oxford superconducting magnet with a liquid helium cryostat. The superconducting magnet is in the form of split pairs with magnetic field vector in the vertical plane of the cryostat. The sample was mounted through the top-loading access along the magnetic field direction. The magnetic field can be tuned from 0 up to 14 T at any temperature between 1.5 K and 300 K. 

The high-pressure NPD measurements were performed using PLANET\cite{Hattori_2015} at J-PARC. The Mito system\cite{Komatsu_2013} equipped with anvils made of ZrO$_2$, was implemented for high-pressure and low-temperature controlling, which can reach highest pressure P = 5 GPa and base temperature T = 77 K. The sample was loaded with a pressure-transmitting medium of deuterated glycerol in a pair of encapsulating TiZr gaskets fitted in a tapered Al ring. The pressure is determined from lattice parameter of Pb based on the EOS\cite{Strassle_2014}. The temperature is measured by two Pt resistance temperature sensors attached to the body of the press. The sample was cooled to 80 K before applying pressure up to $P$ = 1.4 GPa. 
We conducted the symmetry analysis based on the representation theory using the software suits, $BasIreps$\cite{fullprof_1993} and $SARAh$\cite{Wills_2000}, and carried out the Rietveld refinement with the software suites, $Z-Rietveld$\cite{Z-Rietveld_2009, Z-Rietveld_2012} and $FullProf$\cite{fullprof_1993}. 

\subsection{Muon spin relaxation ($\mu$SR)}
Time-differential muon ($\mu^+$) spin relaxation was measured using S-line at J-PARC. The polycrystalline PrBaCo$_2$O$_{5.5+x}$ samples were pressed into pellet of 5 mm thickness and 25 mm in diameter and mounted in a helium-flow cryostat with base temperature $T$ = 4 K. The measurements were conducted under the longitudinal magnetic field up to $H$ = 0.4 T in the parallel direction with respect to the initial $\mu^+$ spin polarization. The data were analyzed using the software suite, $musrfit$\cite{Musrfit_2012}. 

\subsection{Magnetization \& Resistivity}
The isothermal dc-magnetization ($M-H$ curve) were measured by the Quantum Design Physical Property Measurement System (PPMS) at the Cross-Tokai user laboratories. The resistivity under magnetic field were also measured using a standard dc four-probe method on the PPMS. The high-pressure dc-magnetization ($M-T$ curve) was measured up to $P$ = 1.2 GPa using the piston-cylider device implemented on the superconducting quantum interference device (SQUID) magnetometer (MPMS) at Department of Physics, Nihon University. Machine oil was used as the pressure-transmitting medium. The magnetization of high-pressure device was subtracted after the measurement. 

\subsection{Density-functional-theory (DFT) calculations}

DFT calculations were performed using the VASP code\cite{Kresse_1996} with generalized gradient approximation (GGA) potential. In order to account for correlation effects of 3$d$ electrons, we employed the Heyd-Scuseria-Ernzerhof (HSE) screened hybrid functional method\cite{Heyd_2003}, which mixes the exact non-local Fock exchange and the density-functional parametrized exchange. The HSE is known to improve the evaluation of band gap energy and the structural distortion, with respect to GGA+U approaches\cite{Stroppa_2010}. Both the atomic coordinates and the lattice parameters were fully optimized starting with the experimental values while spin$-$orbit interaction is not taken into account. A supercell (with the space group orthorhombic $Pmma$ of ($2a_p \times 2a_p \times 2a_p)$) containing four f.u. of PrBaCo$_2$O$_{5.5+x}$ ($x$ = 0.25) is built to take into account the ordering patterns of oxygen-vacancy and spin/charge/orbital states at eight $Co$ sites as shown in Fig.11a. 

The average valence of $Co$ ion at $x$ = 0.25 is 3.25 so that the eight $Co$ ions in the super cell may show the charge ordering patterns with six trivalent and two quadrivalent ions. Besides, the $Co$ ion may take three spin states in terms of the three configurations of 3$d$ electrons. In the case of trivalent ion, it has low-spin (LS, $t_{2g}^6e_g^0$, $S$ = 0), intermediate-spin (IS, $t_{2g}^5e_g^1$, $S$ = 1) and high-spin (HS, $t_{2g}^4e_g^2$, $S$ = 2) states. This makes the combination number with eight $Co$ charge/spin states enormous in the calculation. To solve the problem, only the spin direction was imposed on each $Co$ 3$d$ state by tuning the density matrix\cite{Yamauchi_2010} while the orbital and charge states were automatically determined when the crystal structure is relaxed. As a result, an antiferromagnetic insulating state was stabilized in an HSE calculation with $E_{gap}$ = 0.2 eV. The ferromagnetic configuration with metallic ground state is also stabilized in the similar procedure.

\section{Results}

At $x$ = 0.24, where the robust NTE and MVE is observed, the ground-state magnetic structure is a G-type antiferromagnetic structure\cite{Miao_PRB_2017} (see the inset of Fig.~\ref{fig1}f). To identify the existence of the ferromagnetic state with smaller volume and slightly higher energy than in the AF state, we investigated simultaneously the nuclear and magnetic structures by neutron powder diffraction (NPD) under hydrostatic pressure and magnetic field.

\subsection{High-pressure NPD}

\begin{figure*}
\includegraphics[width=0.8\textwidth]{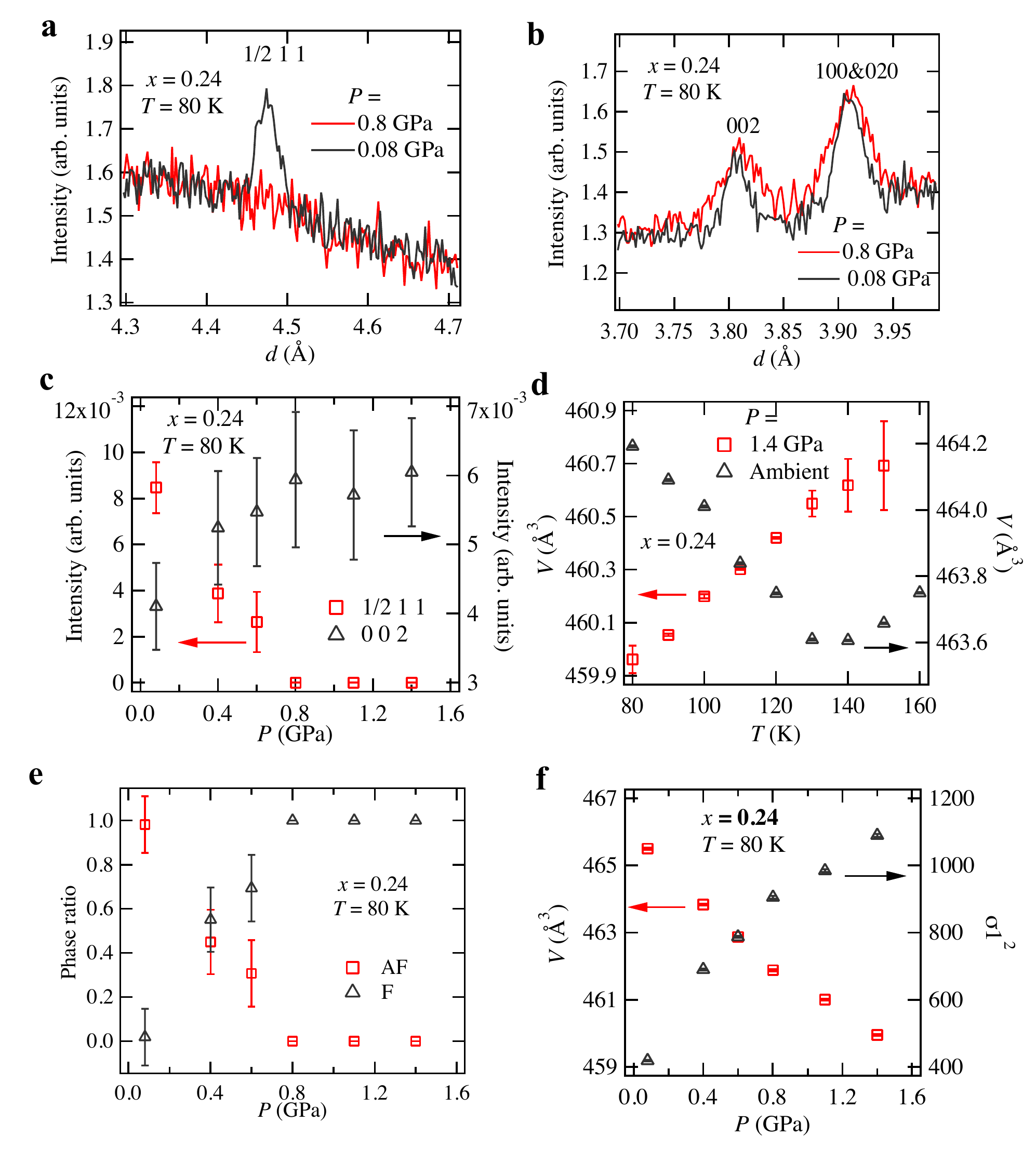}
\caption{\label{fig2} \textbf{$\vert$ Hydrostatic-pressure NPD on PrBaCo$_ \textbf{2}$O$_ \textbf{5.5+x}$ ($ \textbf{x}$ = 0.24) $\vert$} All the indices of reflections are given under the framework of nuclear unit cell ($a_p \times 2a_p \times 2a_p$). There is no indication of symmetry breaking of the crystal structure under hydrostatic pressure up to $P$ = 2.3 GPa. (a, b), Diffraction profiles showing antiferromagnetic (a) and ferromagnetic (b) peaks under ambient pressure and 0.8 GPa at 80 K. In the presence of hydrostatic pressure, the antiferromagnetic reflection $\frac{1}{2}11$ is fully suppressed while the ferromagnetic reflections are identified from the increase in integrated intensity of reflections 002, 100 and 020, indicating the pressure-induced antiferromagnetic$-$ferromagnetic transition. The F peaks become broader upon applying pressure, which is partially due to a non-uniform distribution of pressure in the sample, as confirmed from the peak broadening of Pb used for pressure calibration (see Fig.~\ref{fig3}a and Fig.~\ref{fig3}b). As the pressure increases, the F peaks become even broader than the Pb peaks (see Fig.~\ref{fig3}c and Fig.~\ref{fig3}d), indicating that part of peak broadening in F peaks is inherent to the samples itself. Since the peak broadening does not arise from symmetry breaking (see Fig.~\ref{fig2}f), it might arise from the residual microstrain.  (c), Integrated intensity of the magnetic reflections $\frac{1}{2}11$ and 002 as a function of pressure at 80 K, suggesting the antiferromagnetic?ferromagnetic transition starts from below 0.4 GPa and completes at 0.8 GPa.  (d), Volume of sub unit cell ($2a_p \times 2a_p \times 2a_p)$), obtained from Rietveld refinement, as a function of temperature under ambient pressure and 1.4 GPa, showing the NTE under ambient pressure disappears at high pressure (in the pure F phase). Between 80 to 150 K, the pressure varies from 1.38 GPa to 1.45 GPa. Assuming linear compressibility, the error of volume was estimated by $\sigma=\frac{1.55\%}{2.3GPa}\times[P(T)-1.38GPa]\times460\AA^3$, where the volume decreases by 1.55\% at 2.3 GPa based on the diffraction experiment.(e), Phase fraction for ferromagnetic (F) and antiferromagnetic (AF) phases, obtained from integrated intensity as shown in Fig.~\ref{fig2}c, as a function of hydrostatic pressure at 80 K of $x$ = 0.24. (f), Volume of sub unit cell ($2a_p \times 2a_p \times 2a_p$) and $\sigma_1^2$ of the Gaussian term ($\sigma_G^2= \sigma_0^2+\sigma_1^2d^2+\sigma_2^2d^4$) obtained from Rietveld refinement, as a function of hydrostatic pressure at 80 K of the $x$ = 0.24 sample. The $\sigma_G$ is the Gaussian invariance and Gaussian Half Width at Half Maximum is expressed as $H_G= \sqrt{8ln2}\sigma_G$. The pressure-induced peak broadening can be accounted in by the $\sigma_1^2$ parameter, indicating the peak broadening is isotropic in all peaks and no symmetry lowering of nuclear crystal structure occurs under high pressures.}
\end{figure*}

\begin{figure*}
\includegraphics[width=0.8\textwidth]{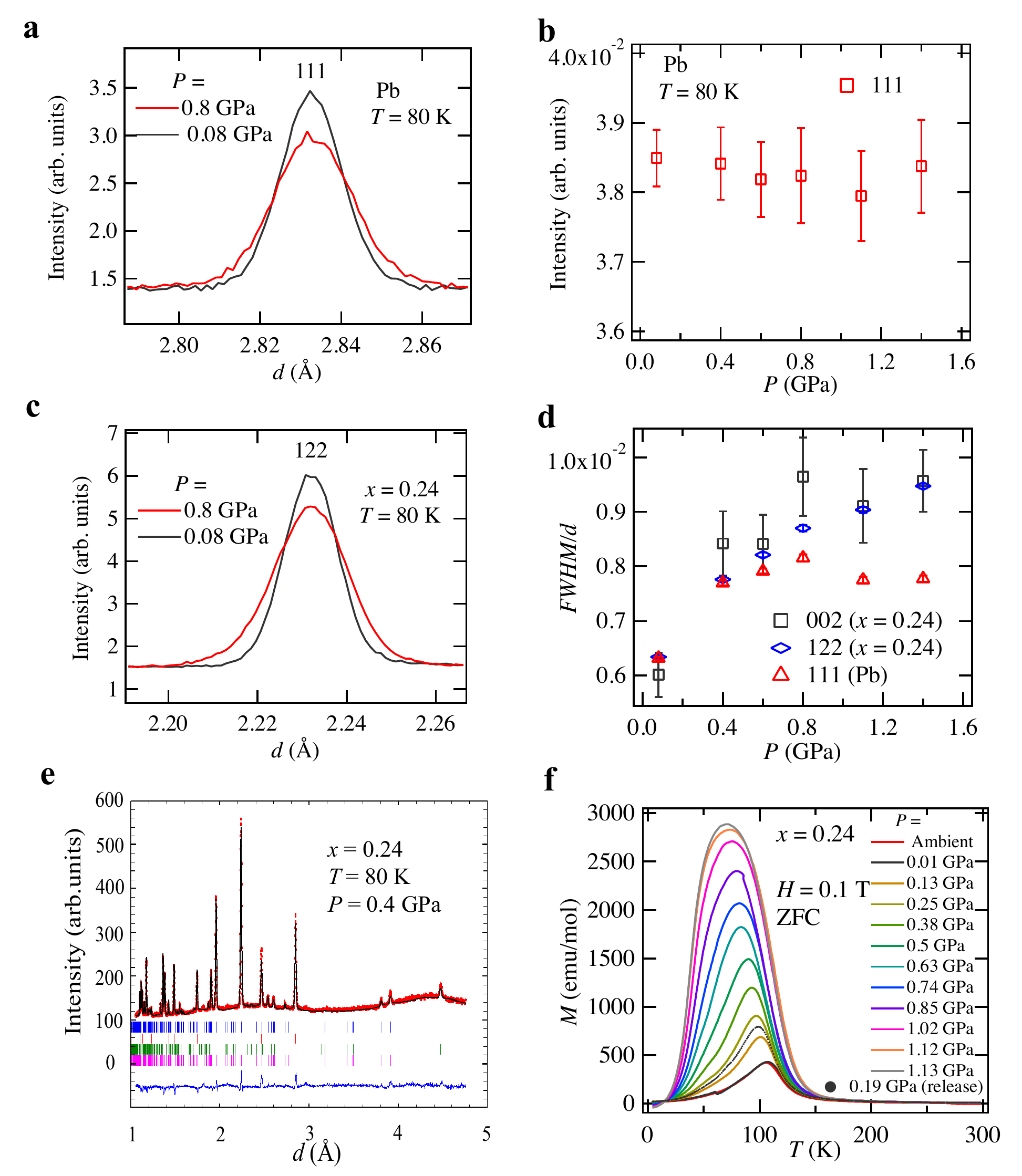}
\caption{\label{fig3} \textbf{$\vert$ High-pressure NPD (a, b) for Pb, high-pressure NPD (c-e) and magnetization (f) for PrBaCo$_ \textbf{2}$O$_ \textbf{5.5+x}$ ($\textbf{x}$ = 0.24) $\vert$} (a), Diffraction profiles of reflection 111 of Pb, which was used for pressure calibration, under 0.08 GPa and 0.8 GPa at 80 K. The Bragg peak becomes much broader with increasing the hydrostatic pressure, indicating that lattice inhomogeneities are induced by a non-uniform distribution of pressure in the sample. (b), Integrated intensity of the reflection 111 of Pb as a function of pressure at 80 K. The independece of pressure corroborates that the peak broadening of Pb in Figure 3b arises from the pressure-induced lattice inhomogenities. Moreover, the comparison of the pressure depenendence of integrated intensity between reflection 111 of Pb and reflection 002 of PrBaCo$_2$O$_{5.5+x}$ ($x$ = 0.24)  in Fig.~\ref{fig2}c, unambiguously reveals that the ferromagnetic reflection grows on top of the reflection 002 in $x$ = 0.24. (c), Diffraction profiles of reflection 122 of $x$ = 0.24, under 0.08 GPa and 0.8 GPa at 80 K. d, Full width at half maximum ($FWHM$) over $d$-spacing $\frac{FWHM}{d}$ (relative peak width) of Bragg reflections (002 and 122 of $x$ = 0.24, and 111 of Pb) as a function of temperature. (g), Rietveld refinement on the data at $T$ = 80 K and $P$ = 0.4 GPa using the model of double phases (AFILV and FMSV). The F and AF structure models are shown in Fig.~\ref{fig1}e. Experimental data points are shown by red dots, and the black line through them is the fit by Rietveld analysis. Since the peaks of high-pressure NPD is too broad to identify the peak splitting from coexistence of the AFILV phase and the FMSV phase, the nuclear structures approximate to the same one as shown by the blue bars. The green bars denote the indices from antiferromagnetic structure and magenta bars represent those of the ferromagnetic structure. The red bars indicate the nuclear structure of Pb. Blue line shows the difference between experiment and calculation. The residual value R$_{wp}$ is 16.30\% and R$_{M}$ is 38.4\%. h, Molar magnetization as a function of temperature (M$-$T curve) measured under magnetic field of 0.1 $T$ through zero-field cooling (ZFC) processes for various hydrostatic pressures. The M$-$T curves at different pressures were measured in the direction of increasing pressure except for that at 0.19 GPa.}
\end{figure*}

Fig.~\ref{fig2} shows the results of high-pressure NPD on the $x$ = 0.24 sample. Melting of AF order is evidenced by the complete suppression of magnetic reflection $\frac{1}{2}11$ from the AF structure at $P$ = 0.8 GPa, $T$ = 80 K (Fig.~\ref{fig2}a), while the formation of F order is observed from the increase in integrated intensity of 002, 100 and 020 (Fig.~\ref{fig2}b)21. Simultaneously, the unit cell volume shrinks by 0.9\% under the same pressure without any symmetry breaking of nuclear crystal structure  (Fig.~\ref{fig2}f). These results lead to the conclusion that the antiferromagnetic large-volume phase transforms to the ferromagnetic small volume phase in the presence of hydrostatic pressure. The pressure dependence of integrated intensity (Fig.~\ref{fig2}c) suggests that the transition initiates from below 0.4 GPa and completes at 0.8 GPa, in broad agreement with results of hydrostatic-pressure magnetization measurement that indicates the onset and end of the transition being 0.13 GPa and 1.12 GPa respectively (Fig.~\ref{fig3}f).

In contrast to the NTE under ambient pressure, the unit cell volume of ferromagnetic small-volume phase at 1.4 GPa exhibits positive thermal expansion (Fig.~\ref{fig2}d), reminiscent of the same behavior in over-doped samples ($e.g.$, $x$ = 0.41 in Fig.~\ref{fig1}a), which has been well described by the Debye-Gruneisen model and shows negligible MVE\cite{Miao_AM_2017}. Both results corroborate that pure ferromagnetic small volume phase exhibits positive thermal expansion. Also, the thermal expansion in pure antiferromagnetic large-volume phase is assumed to be positive in the light of the positive thermal expansion behavior in under-doped samples ($e.g.$, $x$ = 0 in Fig.~\ref{fig1}a). Consequently, the anomalous NTE as well as MVE at $x$ = 0.24 can be attributed to the transition from antiferromagnetic large-volume phase to ferromagnetic small-volume phase (Fig.~\ref{fig1}e).

%Figure2
%

%Figure3

\subsection{High-field NPD}

\begin{figure*}
\includegraphics[width=0.8\textwidth]{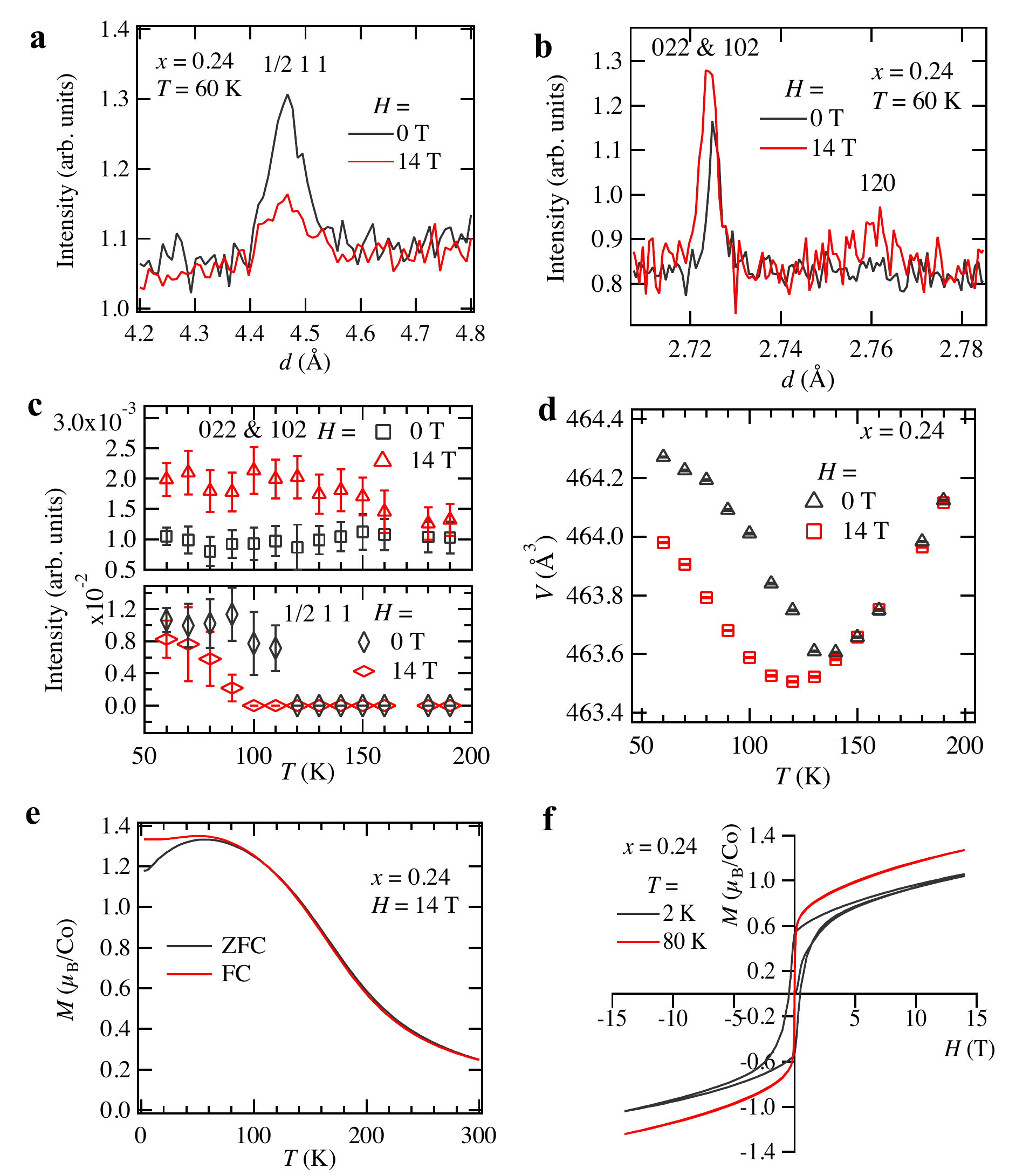}
\caption{\label{fig4}  \textbf{$\vert$ Magnetic-field NPD (a-d) and DC-mageization (e, f) on PrBaCo$_ \textbf{2}$O$_ \textbf{5.5+x}$ ($\textbf{x}$ = 0.24)  $\vert$} All the indices of reflections are given under the framework of nuclear unit cell ($a_p \times 2a_p \times 2a_p$). No sign of symmetry breaking of nuclear crystal structure was observed under magnetic field up to 14 T. (a, b), Diffraction profiles showing antiferromagnetic (a) and ferromagnetic (b) peaks under zero field and 14 T at 60 K. The antiferromagnetic reflection $\frac{1}{2}11$ is suppressed while the ferromagnetic intensity grows on top of nuclear reflections 022, 102 and 120 by applying magnetic field, indicating a field-induced antiferromagnetic$-$ferromagnetic transition. (c) Integrated intensity of the magnetic reflections 022 \&102 (upper panel) and $\frac{1}{2}11$ (lower panel) as a function of temperature under zero field and 14 T respectively, showing that the magnetic field induces ferromagnetic ordering from 150 K while suppressing antiferromagnetic ordering from 120 K upon cooling. (d), Volume of sub unit cell ($2a_p \times 2a_p \times 2a_p$), obtained from Rietveld refinement, as a function of temperature under zero field and 14 T, showing the unit cell significantly contracts with applying magnetic field at low temperatures. (e), Molar magnetization as a function of temperature ($M-T$ curve) measured under magnetic field $H$ = 14 T through both zero-field cooling (ZFC) and field cooling (FC) processes. (f), Molar magnetization as a function of magnetic field ($M-H$ curve) measured at 2 K and 80 K, respectively. }
\end{figure*}

Analogous to hydrostatic pressure, magnetic field can also trigger the transition from antiferromagnetic large-volume phase to ferromagnetic small-volume phase at x = 0.24. As shown in Fig.~\ref{fig4}a and Fig.~\ref{fig4}b, upon applying the magnetic field of 14 T at 60 K, the antiferromagnetic reflection $\frac{1}{2}11$ is significantly suppressed while the ferromagnetic reflections grow on top of the nuclear reflections 022, 102 and 120. From the temperature dependence of integrated intensity (Fig.~\ref{fig4}c), we can see $\frac{1}{2}11$ fully vanishes under the 14 T field at higher temperatures such as 100 K and 110 K, whereas the F reflections initiate from 150 K upon cooling, coincident with the onset of deviation of unit cell volume from that under zero field as shown in Fig.~\ref{fig4}d. Therefore, the contraction of the unit cell is intimately associated with the change of magnetic structure, suggesting a field induced AF-LV to F-SV phase transition. The magnetization data (Fig.~\ref{fig4}e and Fig.~\ref{fig4}f), which shows that F moment reaches saturation below 14 T at 80 K, is in broad consistency with the magnetic-field NPD experiments. The saturation moment of 1.35 $\mu_B$ from the magnetization also agrees well with the zero-field NPD result (Fig.~\ref{fig1}d).

%Figure4

\subsection{High-resolution NPD}

\begin{figure*}
\includegraphics[width=0.8\textwidth]{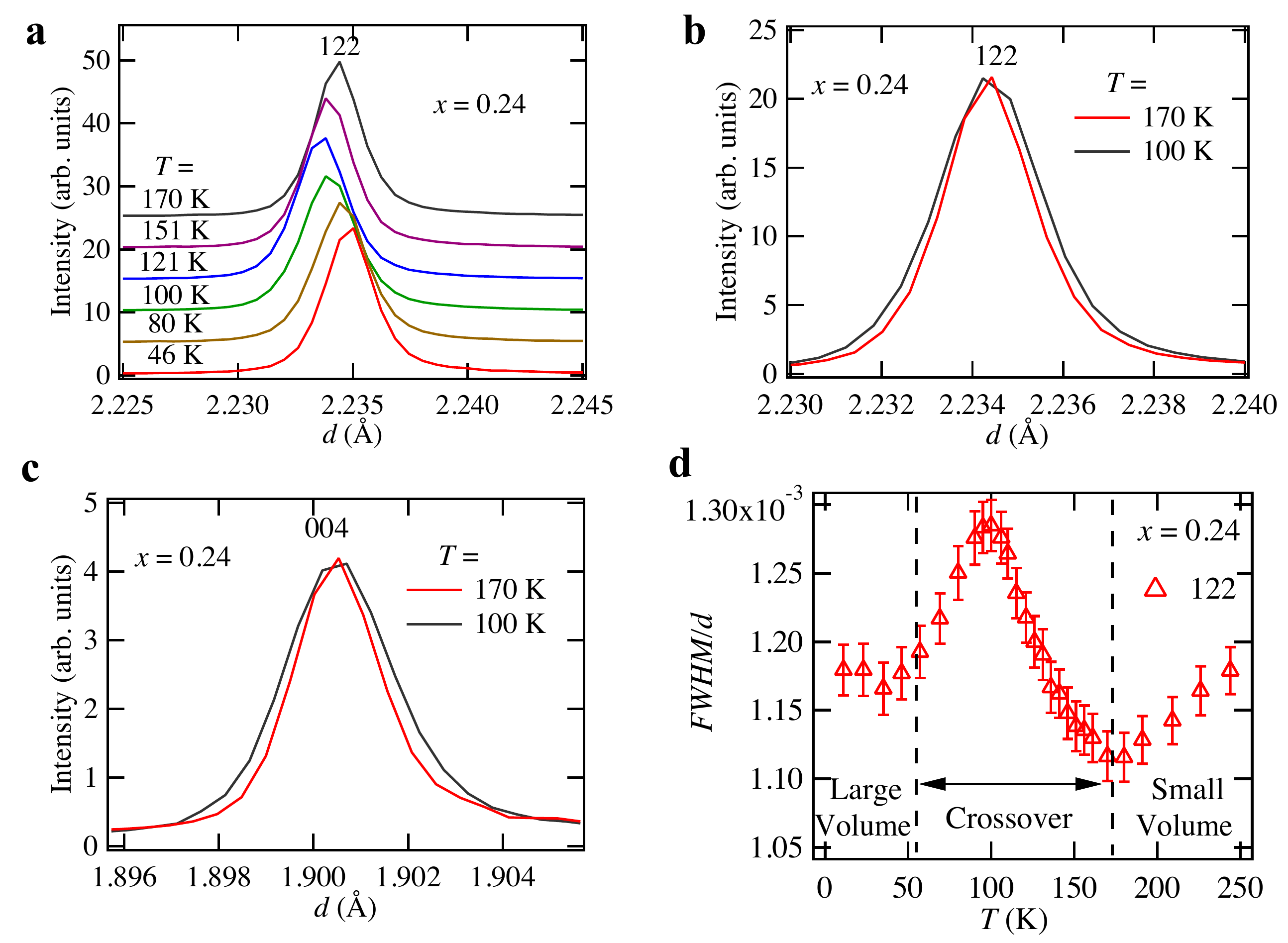}
\caption{\label{fig5}  \textbf{$\vert$ High-resolution NPD on PrBaCo$_ \textbf{2}$O$_ \textbf{5.5+x}$ ($\textbf{x}$ = 0.24)  $\vert$} All the indices of reflections are given under the framework of nuclear unit cell ($a_p \times 2a_p \times 2a_p$). (a), Diffraction profiles of the reflection 122 at different temperatures, showing the single peaks at all temperatures as well as the NTE. (b, c), Comparison of peak width at $T$ = 100 K and $T$ = 170 K for reflection 122 and reflection 004, respectively. The peaks are normalized to the same intensity and shifted to the same position for the better comparison. (d), Full width at half maximum ($FWHM$) over $d$-spacing $\frac{FWHM}{d}$ (relative peak width) of nuclear reflection 122 as a function of temperature. Anomalous peak broadening occurs with decrease in temperature, indicating the transition between FMSV phase and AFILV phase.}
\end{figure*}

Both hydrostatic-pressure and magnetic-field NPD experiments reveal that the antiferromagnetic large-volume phase and ferromagnetic small-volume phase may be the two separate energy minima in the x = 0.24 sample. Therefore, the transition is presumably of discontinuous character, despite that it looks like a continuous phase transition based on the continuous-like $\Delta V$ and $\lvert M \rvert ^2$ (Fig.~\ref{fig1}b). In the case of the discontinuous transition, coexistence of the two phases in the crossover region will bring about lattice inhomogeneities, inducing broadening or splitting of Bragg peak in the diffraction pattern. That is precisely what is measured in the high-resolution NPD experiment on the x = 0.24 sample. As shown by the temperature dependence of relative peak width [full width at half maximum (FWHM) over d-spacing] of nuclear reflection 122 (Fig.~\ref{fig5}d), the peak starts to grow broader as the temperature is decreased to 170 K, reaches maximum width at 100 K and returns to the normal breadth at about 50 K. Moreover, the unusual peak broadening is ubiquitous in all Bragg peaks (Fig.~\ref{fig5}b and Fig.~\ref{fig5}c). We analyzed the shape of all the Bragg peaks with the pseudo-voigt function, a combination of Gaussian and Lorentzian functions, in Rietveld refinement on the diffraction pattern of x = 0.24. The isotropic Gaussian term is expressed as follows: $\sigma_G^2= \sigma_0^2+\sigma_1^2d^2+\sigma_2^2d^4$, where Gaussian Half Width at Half Maximum $H_G= \sqrt{8ln2}\sigma_G$. The isotropic Lorentzian term is as follows: $H_L=\gamma_0+\gamma_1d+\gamma_2 d^2$, where $H_L$ is the Lorentzian Half Width at Half Maximum. We found that the peak width was well described by the isotropic Gaussian and Lorentzian terms and was mainly contributed by the $\sigma_1^2$ and $\gamma_1$  parameters, indicating that the peak broadening comes from lattice inhomogeneities rather than lattice symmetry breaking. The obtained $\sigma_1^2$ and $\gamma_1$ values at 170 K (sharper peaks) are 113(2) $\mu s^2/\AA^2$ and 17.3(2) $\mu s^2/\AA$, respectively, while the two values at 100 K (broader peaks) are 179(2) $\mu s^2/\AA^2$ and 18.0(2) $\mu s^2/\AA$, respectively.

\begin{figure*}
\includegraphics[width=0.8\textwidth]{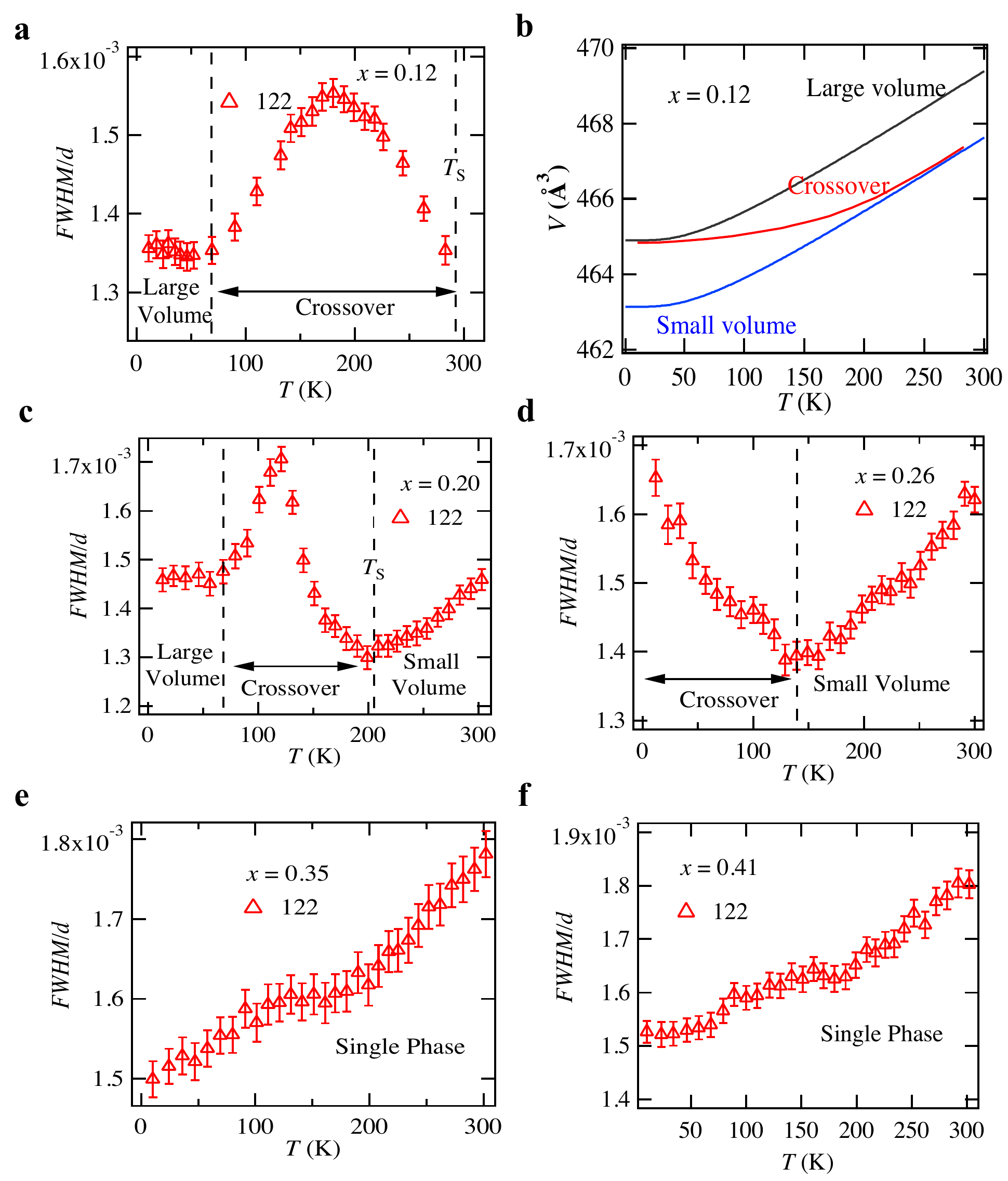}
\caption{\label{fig6} \textbf{$\vert$ High-resolution NPD on PrBaCo$_ \textbf{2}$O$_ \textbf{5.5+x}$  ($\textbf{x}$ = 0.12, 0.20, 0.26, 0.35 and 0.41) and transition model of $\textbf{x}$ = 0.12 $\vert$} All the indices of reflections are given under the framework of nuclear unit cell ($a_p \times 2a_p \times 2a_p$). (a, c, d, e, f), $\frac{FWHM}{d}$ (relative peak width) of nuclear reflection 122 as a function of temperature for hole-doping levels $x$ = 0.12, 0.20, 0.26, 0.35 and 0.41, respectively. (b), The magnetoelectric transition model for NTE of $x$ = 0.12, $i.e.$, the transition from an AFILV phase to a FMSV phase.}
\end{figure*}

On the contrary, the $FWHM$s of those samples that show normal PTE, $e.g.$ $x$ = 0.35 and $x$ = 0.41 (Fig.~\ref{fig6}e and Fig.~\ref{fig6}f), have the monotonic temperature dependence. These results suggest that the unusual peak broadening upon cooling at x = 0.24 corresponds to the coexistence of the LV to SV phases in the critical region. The peak broadening also occurs in other hole doping levels where anomalous thermal expansion occurs, e.g., x = 0.12, 0.20, 0.25 and 0.26 as shown in Fig.~\ref{fig6}. The temperature window could be as broad as from 70 K to 300 K at x = 0.12, giving nearly zero thermal expansion at low temperatures (Fig.~\ref{fig6}b). 

\begin{figure*}
\includegraphics[width=0.8\textwidth]{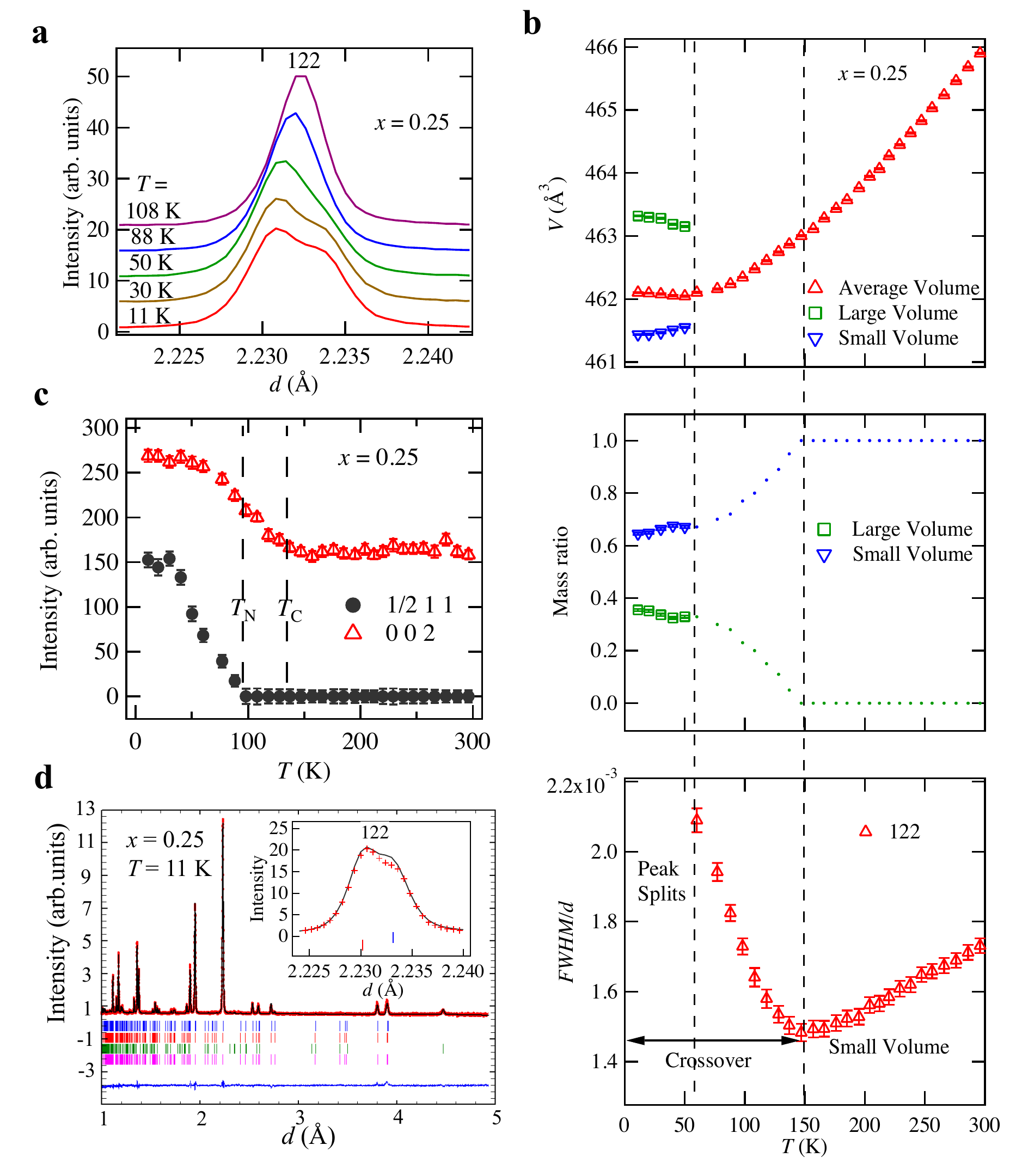}
\caption{\label{fig7} \textbf{$\vert$ High-resolution NPD on PrBaCo$_ \textbf{2}$O$_ \textbf{5.5+x}$ ($\textbf{x}$ = 0.25)  $\vert$}  All the reflection indices are given under the unit-cell framework ($a_p \times 2a_p \times 2a_p$). (a), Diffraction profiles of the reflection 122 at different temperatures. Peak splitting occurs in reflection 122 (and others) at low temperatures, indicating macroscopic phase separation of the AFILV phase and the FMSV phase. (b), Unit cell ($2a_p \times 2a_p \times 2a_p$) volume, mass ratio and $\frac{FWHM}{d}$ of nuclear reflection 122 as a function of temperature.  The $\frac{FWHM}{d}$ of the bottom panel was calculated from the single reflection 122 until it splits into two at about $T$ = 50 K (Fig.~\ref{fig7}a), below which the pattern was described by Rietveld refinement with the model of double phases (Fig.~\ref{fig7}d) , the FMSV phase and the AFILV phase. The resultant volume ($2a_p \times 2a_p \times 2a_p$) and mass ratio as a function of temperature are shown in top and middle panels. The averaged volume is calculated by $V_{average}=R_{AFILV}\times V_{AFILV}+R_{FMSV}\times V_{FMSV}$, where $R$ is the mass ratio. (c), Integrated intensities of antiferromagnetic reflection $\frac{1}{2}11$ and ferromagnetic reflection 002 as a function of temperature, showing the coexistence of antiferromagnetic and ferromagnetic phases. (d), Rietveld refinement on the data at $T$ = 11 K using the model of double phases (FMSV and AFILV). The ferromagnetic and antiferromagnetic structure models are shown in Fig.~\ref{fig1}e. Experimental data points are shown by red dots, and the black line through them is a fit by Rietveld analysis. Red (magenta) bars denote the indices from nuclear (magnetic) structures of FMSV while Blue (green) bars represent the indices from nuclear (magnetic) structures of AFILV phase. Blue line shows the difference between experiment and calculation. The residual values R$_{wp}$ is 5.95\% and R$_{M}$ is 23.0\%.}
\end{figure*}

Further evidence for the two separated energy minima comes from the high-resolution diffraction experiment on the $x$ = 0.25 sample, where macroscopic phase separation was directly probed. As shown in Fig.~\ref{fig7}, the nuclear reflection 122 starts to grow broader upon cooling down to 150 K and completely splits from about 50 K until base temperature, in conjunction with coexistence of the AF reflection $\frac{1}{2}11$ and the F reflection 022 at base temperature. The peak splitting is also observed in many other Bragg peaks, ruling out the possibility of symmetry breaking of nuclear crystal structure. The whole pattern can be well fitted to the combination of the two-phase models (Fig.~\ref{fig7}d), and the resultant volumes and mass ratios as a function of temperature are shown in Fig.~\ref{fig7}b.

%Figure5
%

%Figure6
%

%Figure7

\subsection{Magnetoresistance}

\begin{figure*}
\includegraphics[width=0.7\textwidth]{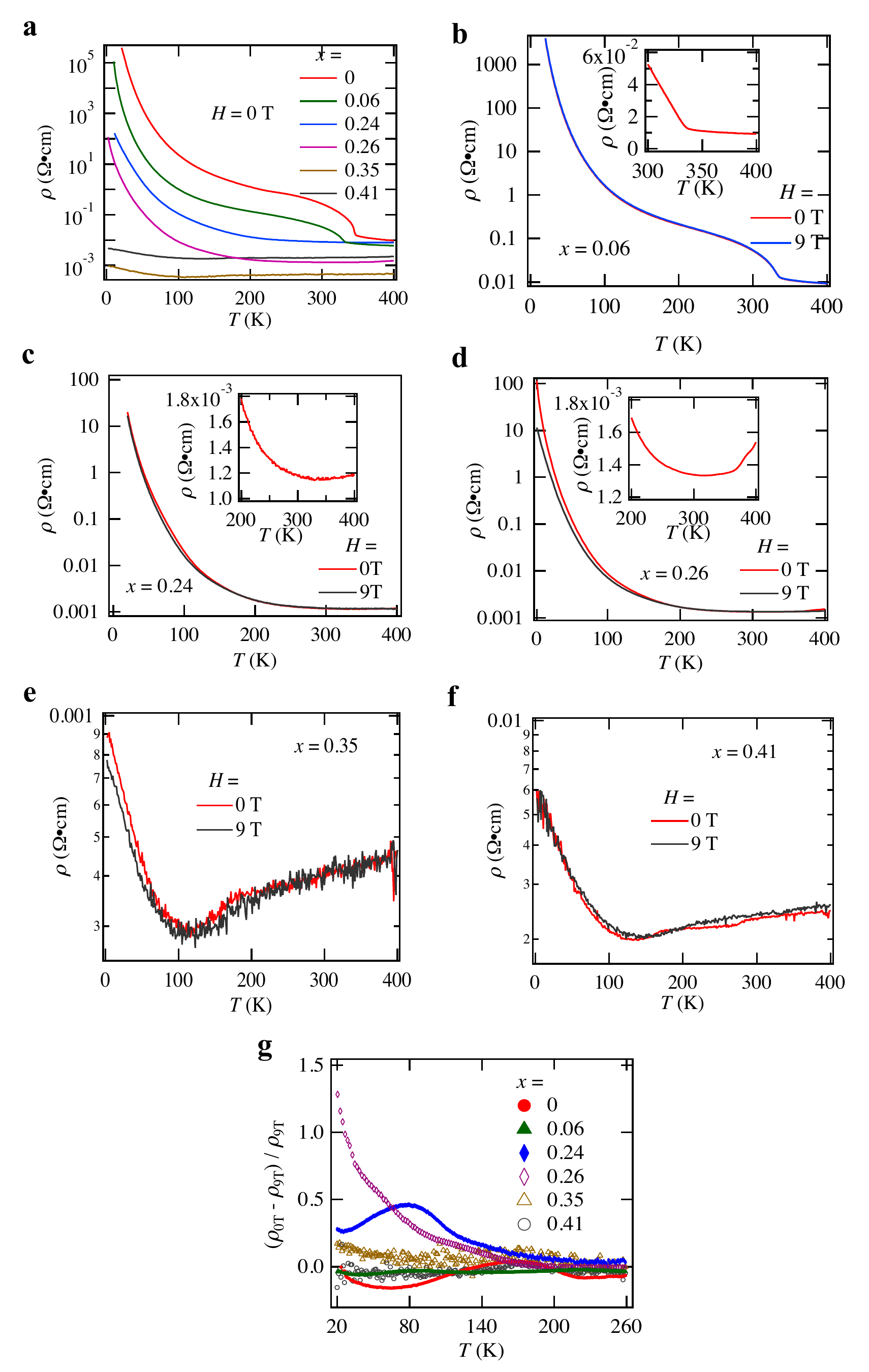}
\caption{\label{fig8} \textbf{$\vert$ Resisitivity measurment of PrBaCo$_ \textbf{2}$O$_ \textbf{5.5+x}$ $\vert$}  (a), Resisitivity as a function of temperature measured under zero magnetic field at various hole doping levels, which shows a sharp decrease in resistivity with increasing hole doping. The relatively lower value in $\rho$ for $x$ = 0.26 and 0.35 than other hole doping $x$ is possibly due to the lower boundary resistance of polycrystal grains compared with that in samples with other hole-doping fractions. (b, c, d, e, f), Resisitivity as a function of temperature measured under zero magnetic field and $H$ = 9 T for $x$ = 0.06, $x$ = 0.24, $x$ = 0.26, $x$ = 0.35 and $x$ = 0.41, respectively. The insets in (b-d) show the resistivity under zero field in a focused temperature window. (g), Magnetoresistance calculated from the relative resistivity difference between $H$ = 0 T and 9 T as a function of temperature at various hole-doping level $x$, which shows that the strongest magnetoresistance occurs near the doping level of the AF$-$F phase boundary.}
\end{figure*}

In addition to the magnetoelastic coupling observed in the diffraction experiment, we also discovered a magnetoelectric coupling from the magnetoresistance measurement. As shown in Fig.~\ref{fig8}, the low-temperature resistivity decreases upon applying magnetic field of 9 T for $x$ = 0.24 and 0.26, whereas the magnetic field barely influence the resistivity at other doping levels away from the antiferromagnetic$-$ferromagnetic boundary in the phase diagram. Such hole-doping dependence is more clearly demonstrated by the MR at low temperatures (Fig.~\ref{fig8}g), which reaches maximum near the antiferromagnetic$-$ferromagnetic boundary. Taking into account the fact that the doping dependence of MVE also shows a peak centered near the antiferromagnetic$-$ferromagnetic phase boundary (Fig.~\ref{fig1}c), the coincidence of MVE and magnetoresistance indicates a strong interplay among the properties of lattice, magnetism and electronic transport, which induces the competing AFILV and FMSV ground states.

%Figure8

\subsection{$\mu$SR}

\begin{figure*}
\includegraphics[width=0.7\textwidth]{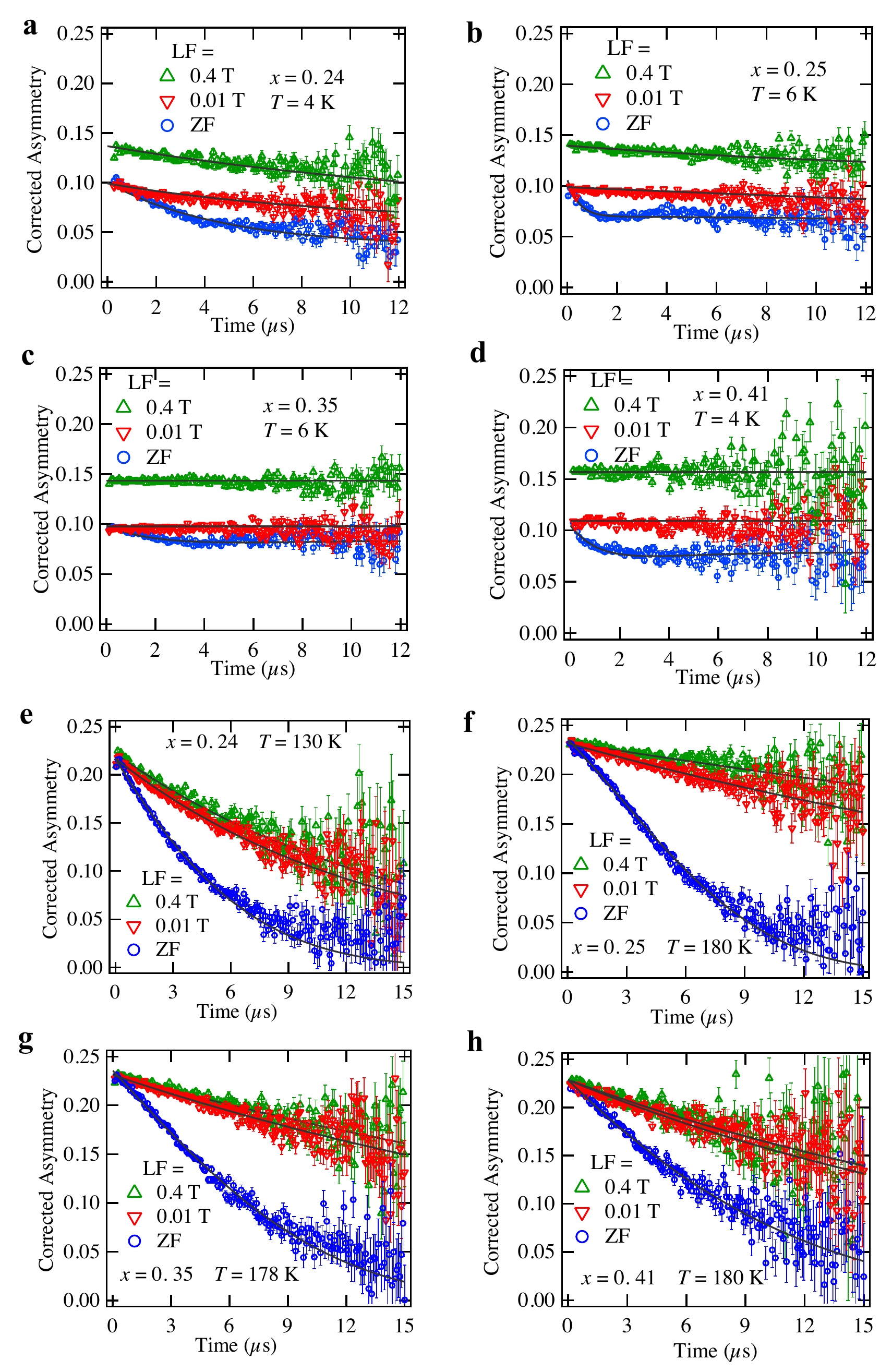}
\caption{\label{fig9} \textbf{$\vert$ Muon spin relaxation ($\mu$SR) measurements of PrBaCo$_ \textbf{2}$O$_ \textbf{5.5+x}$ ($\textbf{x}$ = 0.24, 0.25, 0.35 and 0.41) $\vert$}  (a, b, c, d), $\mu$SR spectra of $x$ = 0.24 at $T$ = 4 K (a), $x$ = 0.25 at 6 K (b), $x$ = 0.35 at 6 K (c) and $x$ = 0.41 at 4 K (d) under various longitudinal fields (LFs). The spectrum of zero field at long $t$ is significantly lifted up with application of a weak LF 0.01 T, which arises from the decouplig of muon from the nuclear moments or spin glass component. The fitting equation for describing the zero-filed (ZF) spectrum is $A_sG_z(H, t) = G_{GKT}(t)\{A_1exp[-\Lambda t] + A_2[\frac{1}{3} + \frac{2}{3}cos(\gamma_{\mu}Bt)]\}$ for $x$ = 0.24, 0.25 and 0.35, $A_sG_z(H, t) = G_{LKT}(t)\{A_1exp[-\Lambda t] + A_2[\frac{1}{3} + \frac{2}{3}cos(\gamma_{\mu}Bt)]\}$ for x = 0.41, where $G_{GKT}(t)$ is the Gaussian Kubo$-$Toyabe relaxation function (nuclear moments distribution),  $G_{LKT}(t)$  the Lorentzian Kubo$-$Toyabe relaxation function (spin glass component), and the other terms are described in the main text for Equation~\ref{eqn:equation1}. In all samples, the spectra under LFs are fitted to Equation~\ref{eqn:equation1} because the nuclear moment component [$G_{GKT}(t)$] and spin glass component [ $G_{LKT}(t)$ ] are fully decoupled by LFs. All the fittings are shown as the solid lines. (e, f, g, h), $\mu$SR spectra of $x$ = 0.24 at $T$ = 130 K (e), $x$ = 0.25 at 180 K (f), $x$ = 0.35 at 178 K (g) and $x$ = 0.41 at 180 K (h) under various longitudinal fields (LFs). The zero-filed (ZF) spectra indicate that the background is negligible. Since all spectra were collected above magnetic ordering temperatures, only the dynamic spin fluctuation and nuclear spin distribution were taken into account and all the spectra are fitted to the equation $A_sG_z(H, t) = A_1exp[-\Lambda t]G_{GKT}(t)$, where $G_{GKT}(t)$ amounts to 0 under LFs since $G_{GKT}(t)$ is fully decoupled by LFs. The solid lines show the fittings.}
\end{figure*}

Since the crossover between the two energy minima is easily activated with small amount of external energy (pressure, magnetic field $etc.$), strong phase fluctuations are anticipated near the boundary at $x\approx$0.24, for which we obtained direct evidence from muon-spin-relaxation ($\mu$SR) measurement. As shown in Fig.~\ref{fig9}a and Fig.~\ref{fig9}b, dynamic fluctuations of $Co$ spins at $x$ = 0.24 and 0.25 at base temperatures $T$ = 4 K and 6 K are identified from the decay of spectra under longitudinal field (LF) up to $H$ = 0.4 T. On the contrary, as shown in Fig.~\ref{fig9}c and Fig.~\ref{fig9}d, the lack of time dependence in the spectra of the $x$ = 0.35 and $x$ = 0.41 samples even under the weak LF $H$ = 0.01 T suggests that all the $C$o spins are static at base temperatures. The difference unambiguously illustrates that the spin fluctuations, arising from the AFILV$-$FMSV phase fluctuations, become much stronger in the vicinity of the phase boundary than being away from it. Since the time window of $\mu$SR measurement is on the scale of $\sim$$\mu$s, the observed phase fluctuation rate falls into the range of $\sim10^6$ Hz.

%Figure9

As shown in Fig.~\ref{fig9}, the reduction in initial asymmetry $A_s$ (at $t$ = 0) at zero field upon decreasing from high temperatures to base temperatures and the increase in $A_s$ with applying LFs at base temperatures both indicate that all the samples contain a static phase with internal field at low temperatures, corresponding to the long-range $Co$-spin orders that were detected by NPD\cite{Miao_AM_2017}. The magnetic volume fractions for both dynamic and static phases can be derived by taking the advantage of the spectra under the $H$ = 0.01 T, which decouples the decay from nuclear moments with least disturbance to the contributions from $Co$-moments. Since the background is negligible according to the high-temperature spectra [Fig.~\ref{fig9}(e-g)], the spectra under $H$ = 0.01 T can be fitted to the following equation:

%Equation1
\begin{eqnarray}
 A_sG_z(H, t) = A_1exp[-\Lambda t] + A_2[\frac{1}{3} + \frac{2}{3}cos(\gamma_{\mu}Bt)]
\label{eqn:equation1}
\end{eqnarray}

\begin{figure*}
\includegraphics[width=0.8\textwidth]{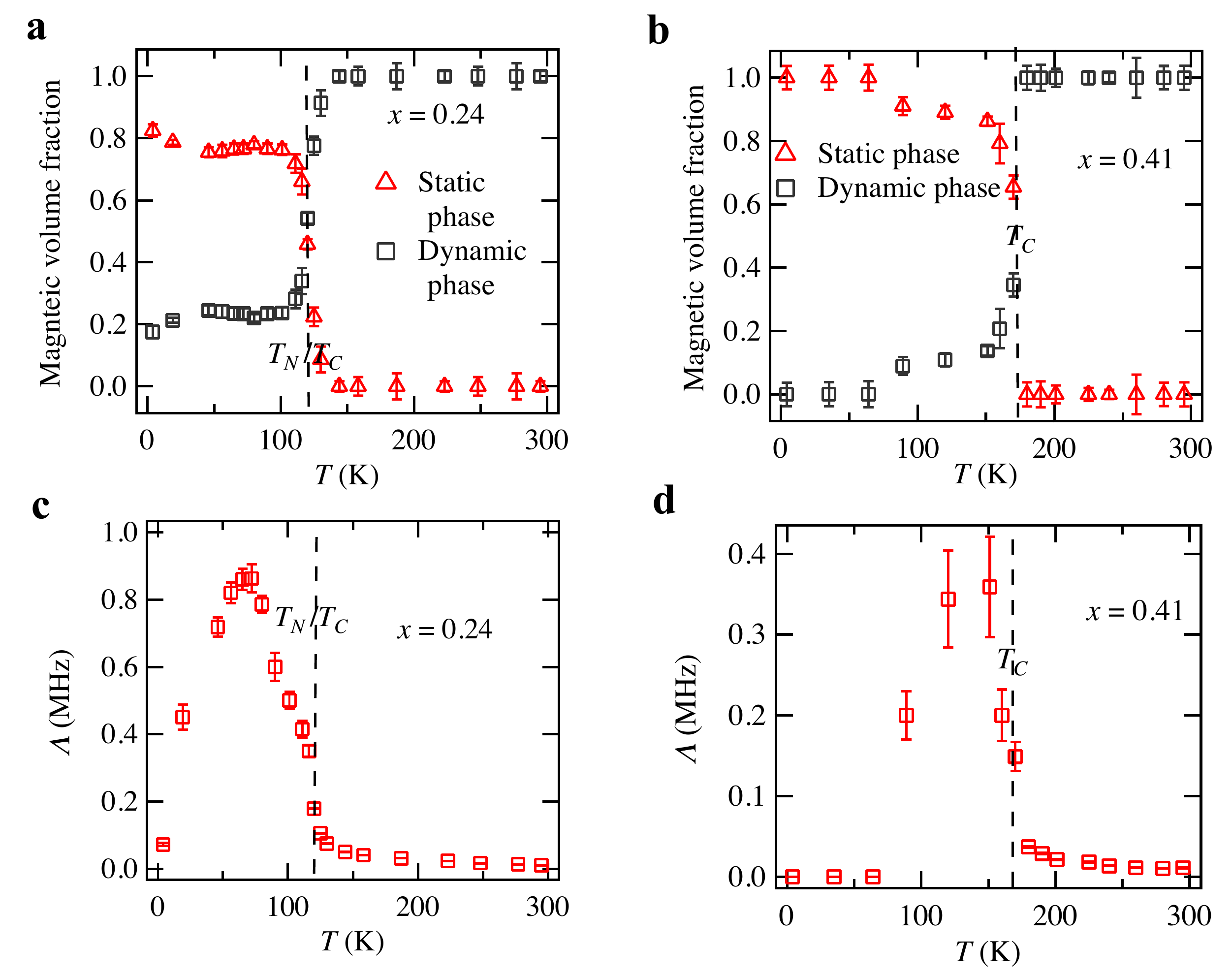}
\caption{\label{fig10} \textbf{$\vert$ Muon spin relaxation ($\mu$SR) measurements of PrBaCo$_ \textbf{2}$O$_ \textbf{5.5+x}$ ($\textbf{x}$ = 0.24, and 0.41) $\vert$}  (a, b), Magnetic volume fractions as a function of temperature for $x$ = 0.24 (a) and $x$ = 0.41 (b), derived from Fraction (dynamic phase) = $\frac{A_1}{A_1+ A_2}$ and Fraction (static phase) =  $\frac{A_2}{A_1+ A_2}$, where $A_1$ and $A_2$ are obtained from the fittings of the spectra under LF of 0.01 T to Equation~\ref{eqn:equation1}. c, d, Temperature dependence of muon spin-lattice-relaxation rate $\Lambda$ for $x$ = 0.24 (e) and $x$ = 0.41 (f), derived from fittings of the spectra under LF of 0.01 T to Equation~\ref{eqn:equation1}. The $\Lambda$ for $x$ = 0.41 reduces to 0 at about upon cooling down to about 80 K, which is commonly observed in those materials showing magnetic ordering. However, the $\Lambda$ for $x$ = 0.24 shows unusual behavior, $i.e.$, it stays nonzero even down to the base temperature. The results imply that the giant phase fluctuations occur at $x$ = 0.24 so that the related dynamic spin fluctuations survive until the base temperature.}
\end{figure*}

Where $A_1$ and $A_2$ parameterize the contributions from dynamic and static phases, respectively. The oscillation couldn't be observed due to the limitation of the time resolution of pulsed muon source and the term $cos(\gamma_{\mu}Bt)$ is averaged to 0 here. The temperature dependence of magnetic volume fractions calculated from  $A_1$ and $A_2$, and the damping rate $\Lambda$ (Fig.~\ref{fig10}) reveals that the Co spins at $x$ = 0.41 become completely static as the temperature is decreased to 80 K whereas the spin fluctuations at $x$ = 0.24 survives until the base temperature, which again corroborates the strong phase fluctuation near the phase boundary.

%Figure10

\subsection{DFT calculations}

\begin{figure*}
\includegraphics[width=0.8\textwidth]{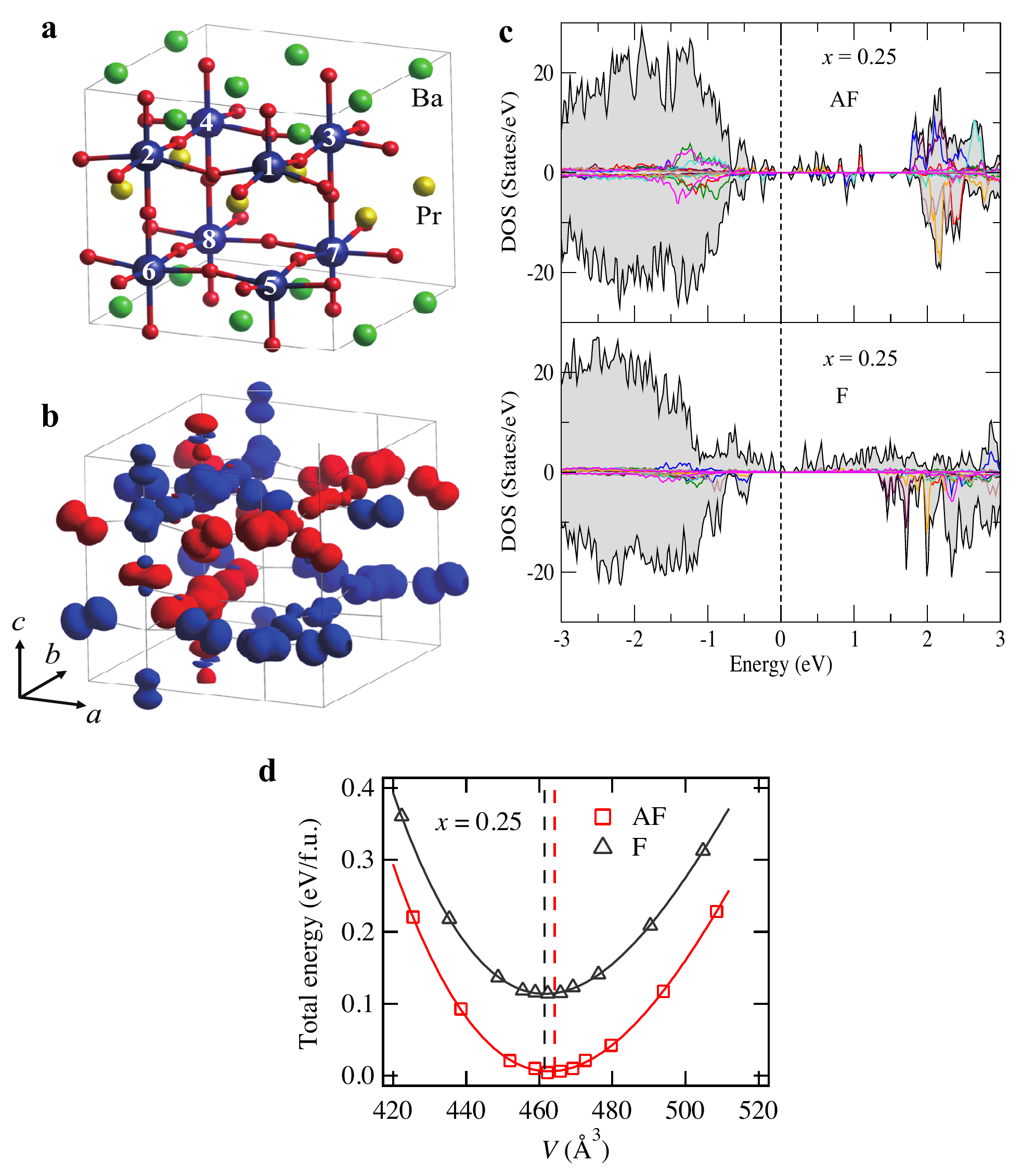}
\caption{\label{fig11} \textbf{$\vert$ DFT calculations of PrBaCo$_ \textbf{2}$O$_ \textbf{5.5+x}$ ($\textbf{x}$ = 0.25) $\vert$}  (a), Sub unit cell  ($2a_p \times 2a_p \times 2a_p$) used in DFT calculations. Numbers indicate $Co$ site index. (b), Spin density (blue: up-spin and red: down-spin state) of Co-3d orbital states in an energy range of $E_F - 0.5eV< E < E_F$ in the antiferromagnetic (AF) state. (c), Density of states for antiferromagnetic (AF, upper panel) and ferromagnetic (F, lower panel) states of $x$ = 0.25, showing the antiferromagnetic state has an open gap while the ferromagnetic phase is gapless. The dashed line denotes Fermi level. The black shadowed lines show the total density of states and the colored lines indicate the contribution from each of the eight Co sites in the nuclear subrcell  ($2a_p \times 2a_p \times 2a_p$). (d), Calculated total energies for the AF and F states as a function of cell volume, fitted to a polynomial function to derive the energy minima as shown by dashed lines.}
\end{figure*}

We also found theoretical support for the AFILV$-$FMSV transition scenario. As shown in Fig.~\ref{fig11}c, the density-functional-theory (DFT) calculations on $x$ = 0.25 demonstrate that an electronic gap is open at the Fermi level of the antiferromagnetic state while the Fermi level in ferromagnetic state become gapless. The density of states at Fermi level of the ferromagnetic state is quite low, which is consistent with our experimental observation that the metallic behavior can be easily destroyed by the AFILV clusters. The volume dependence of the total energy in Fig.~\ref{fig11}d shows separate minima of antiferromagnetic and ferromagnetic state respectively, where the unit cell volume of the energy minimum of antiferromagnetic state 463.74 $\AA^3$ and that of the ferromagnetic state is 460.94 $\AA^3$. Therefore, the AFILV and FMSV ground states are successfully reproduced by the DFT calculations.

%Figure11

\section{Discussion}

With compelling evidences from multiple techniques, we unveil the microscopic origin for the volume-magnetism correlation in PrBaCo$_2$O$_{5.5+x}$. Specifically, giant competition between the AFILV phase and the FMSV phase occurs in the vicinity of the phase boundary, so that dramatic magnetoelastic and magnetoelectric responses can be driven by moderate external stimuli such as temperature, hole doping, hydrostatic pressure and magnetic field etc. As a result, the anomalous NTE (Fig.~\ref{fig1}a) originate from the temperature-induced AFILV$-$FMSV transition. The strong competition between AFILV and FMSV ground states manifest itself in the NPD, magnetoresistance, and $\mu$SR experiments as well as the DFT calculations. Actually, the temperature dependence of resistivity under zero field also exhibit such competition. As shown in the inset of Fig.~\ref{fig8}c and Fig.~\ref{fig8}c, the metallic behavior is found at temperatures over 300 K and gradually transformed into the insulting behavior at low temperatures for both $x$ = 0.24 and 0.26, indicating the transition from FMSV phase to AFILV phase upon cooling. The transition temperature observed from the resistivity measurement is higher than that from the diffraction measurement, which is presumably due to that the insensitivity of NPD to the short-range ordered AFILV clusters at high temperatures. Similarly, the insulting behavior at low temperatures in x = 0.35 and 0.41 (Fig.~\ref{fig8}e and Fig.~\ref{fig8}f) also comes from the AFILV clusters embedded in FMSV matrix, which is evidenced from the spin glass behavior observed in both ac-magnetization\cite{Miao_AM_2017} and zero-field $\mu$SR measurement (Fig.~\ref{fig9}d).

The coexistence of two phases in the anomalous thermal expansion regions of the samples ($x$ = 0.12, 0.20, 0.24, 0.25 and 0.26) are evidenced from ubiquitous broadening of Bragg peaks upon cooling in the high-resolution NPD experiment. The peak broadening might develop into peak splitting when the relatively peak width $\frac{FWHM}{d}$ is over 0.20\% within the present instrument resolution, which was observed in the $x$ = 0.25 sample. Accordingly, the AFILF$-$FMSV transition is intrinsic of discontinuous character, despite that it looks like a continuous phase transition from the volumetric and magnetic order parameters\cite{Miao_AM_2017}. Weak discontinuous phase transitions were often characterized as continuous under the normal experimental precision and the conclusion can be overturned by higher-precision measurements\cite{Yang_2008}. PrBaCo$_2$O$_{5.5+x}$ constitutes such case as well and the merit of high resolution allows us to finally determine the correct transition type.

In the AFILV phase, super-exchange antiferromagnetic interaction dominates magnetic correlation between the $Co$ ions, leading a tendency towards the insulating behavior, and with additional help from longer atomic distance, the insulating ground state is stabilized. As for the FMSV phase, the $Co$ ions are mainly correlated by double-exchange ferromagnetic interaction which prefers to the itinerant charge transport, and the shorter atomic distance further helps to stabilize the metallic ground state. To be noted, the volume difference between AFILV and FMSV phases is not related with change of spin state of $Co^3+$ ions, $i.e.$, the size of magnetic moment, since experimentally we did not observe nominal change in the magnetic moment size at 10 K across the boundary of the two phases (Fig.~\ref{fig1}d). Compared with most of MVE materials like Invar alloys\cite{Wasserman_1990, Moriya_1980}, trivalent manganese fluoridecite\cite{Hunter_2004}, manganites\cite{Garcia_1997}, antiperovskite manganese nitrides\cite{Takenaka_2005, Guo_2015, Deng_2015}, intermetallics\cite{Hu_PRA_2019, Song_2018, Hu_IC_2019} $etc.$, where the MVE originates from the coupling between spin and lattice degrees of freedom, PrBaCo$_2$O$_{5.5+x}$ exhibits an unusual volume$-$magnetism correlation where the spin, charge and lattice degrees of freedom are all intimately connected so that MVE and ME occur simultaneously. Our study unveils a new mechanism for the MVE and opens an alternative path to the design of MVE materials. 

Among the ME materials, such as CMR or multiferroic compounds, PrBaCo$_2$O$_{5.5+x}$ is a rare example that the symmetry of the crystal structure survives in the AFILV$-$FMSV transition. In manganites with CMR effect, charge and/or orbital ordering stabilize the antiferromagnetic insulating phase. Charge ordering brings about the loss of translation symmetry and orbital ordering couples with Jahn-Teller distortion so that the antiferromagnetic insulating phase resides in a low-symmetry crystal structure.  Applying magnetic field can melt the orders and transform it into a high-symmetry ferromagnetic metallic phase\cite{Tokura_2006, Dagotto_2001}. In multiferroics, the magnetic ordering causes through inverse Dzyaloshinksii$-$Moriya interaction the structural distortion, which breaks the inversion symmetry and induces the electronic polarization\cite{Khomskii_2009, Cheong_2007}. The ME in PrBaCo$_2$O$_{5.5+x}$ does not require either charge/orbital ordering or inversion symmetry breaking because the charge transport property can be significantly influenced by the unit cell volume. Therefore, our study here demonstrates a different way of generating ME.

The easy AFILV$-$FMSV phase conversion gives the cobaltite broad tunability of average volume via multiple external stimuli. The broad tunability can be utilized to realizing zero thermal expansion in a wide temperature window, which is of great importance for industrial use\cite{Chen_2015, Barrera_2005}. For example, reducing the doping level $x$ from 0.24 to 0.12 opens a wider temperature window of transition (50 - 170 K at $x$ = 0.24, see Fig.~\ref{fig5}d; 70 - 300 K at $x$ = 0.12, see Fig.~\ref{fig6}a), and transforms the NTE into the nearly zero thermal expansion. What could be more interesting is that moderate magnetic field, hydrostatic pressure, or combination of both will simultaneously produce multiple responses from the lattice, magnetism and charge transport properties. The multiple responses are promising for new technical applications like magnetic/pressure sensors, actuators, transducers and so on.

\begin{acknowledgments}
We acknowledge the merit award of beam time for neutron and muon experiments at Japan Proton Accelerator Research Complex (J-PARC). The neutron scattering experiment was approved by the Neutron Science Proposal Review Committee of J-PARC/MLF (Proposal No. 2014S05 for SuperHRPD, No. 2017B0062 for PLANET, No. 2017A0014 and 2017B0045 for S-line) and supported by the Inter-University Research Program on Neutron Scattering launched by Institute of Materials Structure Science, High Energy Accelerator Research Organization. We thank Dr. Yukio Noda, Dr. Masatosh Hiraishi and Dr. Soshi Takeshita, Dr. Yang Ren, Dr. Jiaxin Zheng, Dr. Mouyi Wen and Mr. Zongxiang Hu for helpful discussions. We also appreciate Dr. Hirotaka Okabe, Dr. Motoyuki Ishikado, Dr. Taketo Moyoshi, Dr. Masato Hagihara, Mr. Masahiro Shioya, Ms. Widya Rika and Ms. Nur Ayu for their assistance with the experiments and data analysis.
\end{acknowledgments}

\bibliography{Reference}

\end{document}